# Non-LTE models for synthetic spectra of type Ia supernovae

## IV. A modified Feautrier scheme for opacity-sampled pseudo-continua at high expansion velocities and application to synthetic SN Ia spectra

T. L. Hoffmann[1], D. N. Sauer[2], A. W. A. Pauldrach[1], and P. J. N. Hultzsch[1]

[1] Universitätssternwarte München, Scheinerstr. 1, D-81679 München, Germany
   e-mail: hoffmann@usm.uni-muenchen.de; uh10107@usm.uni-muenchen.de; pjnh@usm.lmu.de
[2] Meteorologisches Institut, Ludwig-Maximilians-Universität München, Theresienstr. 37, D-80333 München, Germany
   e-mail: daniel.sauer@lmu.de

**ABSTRACT**

*Context.* Type Ia supernovae (SN Ia) have become an invaluable cosmological tool as their exceptional brightness makes them observable even at very large distances (up to redshifts around $z \approx 1$). To investigate possible systematic differences between local and distant SN Ia requires detailed models whose synthetic spectra can be compared to observations, and in which the solution of the radiative transfer is a key ingredient. One commonly employed method is the Feautrier scheme, which is generally very robust, but which can yield wrong results under certain conditions that frequently occur in the modelling of supernova ejecta or even in the radiatively driven expanding atmospheres of hot stars.
*Aims.* We attempt to devise an improvement of the procedure we have developed for simulating the radiative transfer of metal-rich, intermediate- and low-density, line-dominated atmospheres and that allows the method to be applied successfully even under conditions of high expansion velocities.
*Methods.* We use a sophisticated model atmosphere code which takes into account the non-LTE effects and high velocity gradients that strongly affect the physics of SN Ia atmospheres at all wavelengths to simulate the formation of SN Ia spectra by the thousands of strong spectral lines which intricately interact with the "pseudo-continuum" formed entirely by these Doppler-shifted lines themselves. We focus to an investigation of the behavior of the Feautrier scheme under these conditions.
*Results.* Synthetic spectra of SN Ia, a complex product of computer models replicating numerous physical processes that determine the conditions of matter and radiation in the ejecta, are affected by large spatial jumps of the line-dominated opacities and source functions for which the application of even well-established methods may harbor certain pitfalls. We analyze the conditions that can lead to a breakdown of conventional procedures and we derive for the Feautrier radiative transfer solver a modified description which yields more accurate results in the given circumstances.

## 1. Introduction

Devising a tool that can relate the observed characteristics of SN Ia to specifics of the underlying explosion is of fundamental interest not only for the study of the explosion mechanism itself (which is still unclear in many details) but also for investigating possible systematic effects in the evolution of SN Ia from earlier cosmological times to recent times. The spectra in particular offer a wealth of information that may be utilized for this purpose, but due to the complexity of the problem as a result of back-reactions and the large number of variables involved in the formation of the spectrum the interpretation of the spectra must rely on computer models that replicate as closely as possible the physical processes that determine the conditions of matter and radiation in the ejecta. Realistic radiative transfer models thus provide the crucial link between explosion models and observations.

For transfer problems in general, and radiative transfer in particular (see, e.g., Cannon 1970), Feautrier's (1964) solution to the radiation transfer equation has proved to be enormously successful in many areas of astrophysics. The reasons for its success are that it requires comparatively little computational effort, is numerically stable, and provides good accuracy (cf. Kalkofen & Wehrse 1982).

In our model atmosphere code *WM-basic* (Pauldrach et al. 1998, 2001; Pauldrach 2003; Pauldrach et al. 2012), we employ the Feautrier method to solve the radiative transfer in the first part of a two-stage iteration procedure. In this first part, we use an approximate but fast opacity sampling method on a comparatively coarse radial grid to account for the influence of spectral lines on the radiation field, while in the second part we use a detailed formal integral on a spatial microgrid which is capable of resolving the details of the line profiles and correctly treats all multi-line effects. The closer the results of the first, approximate, method are to those of the detailed, exact method, the less iterations of the second, numerically much more expensive method are required.

Although this procedure has proven to be very reliable in stellar atmospheres for analyzing spectra of expanding atmospheres of hot stars (cf. Pauldrach 1987; Pauldrach et al. 1990; Pauldrach et al. 1994; Pauldrach et al. 2001; Pauldrach et al. 2004; Kaschinski et al. 2012; Pauldrach et al. 2012; Kaschinski et al. 2013) and gaseous nebulae (cf. Hoffmann et al. 2012 and Weber et al. 2013) the physical conditions in SN Ia con-stitute a stronger challenge to the method. In SN Ia the dominating source of opacity are lines, while the continuum plays only a minor role (cf. Pauldrach et al. 1996). Here one often encounters the situation that single lines dominate the opacities by several orders of magnitude, and due to the large velocity gradients in the SN Ia ejecta (homologous expansion, $v \propto r$) these lines are strongly Doppler-shifted as well. Because of the usually not very high radial resolution used in the iterations employing the sampling procedure, it frequently happens that at a particular wavelength point the radial dependencies of the opacities and the source functions show large jumps as the line gets Doppler-shifted through the wavelength grid in the observer's frame. As we show in Section 2, the standard Feautrier solution in the form it is used here, fails to correctly describe the radiation field in the presence of large radial opacity jumps such as those occurring in the metal-rich expansion dominated ejecta of supernovae. In Section 3 we describe an extension that cures this failure and provides a better description of the radiation field without significantly increasing the complexity of the radiative transfer solver. Results of test calculations are presented in Section 4, in Section 5 we apply our new approach in the actual transport code to calculate improved synthetic spectra for SN Ia, and in Section 6 we present the implications of our findings.

## 2. Concept for realistic atmospheric models of SN Ia

To determine observable physical properties of supernovae via quantitative spectroscopy the radiative transfer models must be setup in some detail, and, therefore, detailed physics must already implemented to calculate synthetic spectra on basis of stationary and spherically symmetric envelopes with "photospheres" as lower boundaries precisely. In order to simulate and understand the processes involved in SN Ia envelopes different numerical approaches have been developed to model the radiative transfer and the evolution of the light curves.[1]

In our approach the description of radiative processes in the supernova ejecta is divided into two parts: First, the deposition of radiative energy via $\gamma$-photons resulting from the decay of $^{56}$Ni and $^{56}$Co, as well as the trapping of photons in the highly opaque, expanding material is treated in a time-dependent simulation using a Monte-Carlo approach (Pauldrach et al. 2013). (It is the interplay between deposition of energy and trapping and escape of photons which leads to the characteristic shape of the light curve.) Second, the actual formation of the spectrum is treated with a detailed non-LTE spectral synthesis code. Since this needs to consider only the outer parts of the ejecta from which radiation actually emerges and where the timescales for interaction of photons with matter become small, this problem can be treated in the form of "snapshots" for a given epoch.

Our focus lies on a sophisticated description of the SN Ia spectra with emphasis on high spectral resolution and a detailed modelling of the emission and absorption processes leading to these spectra, in order to quantify the observable physical properties of SN Ia accurately. The spectral synthesis code must thus provide a consistent solution of the non-LTE occupation numbers of all relevant elements (via a solution of the statistical equilibrium based on accurate atomic models, and including collisional and radiative transition rates for all important ions, low-temperature dielectronic recombination, and Auger ionization due to K-shell absorption of soft X-ray radiation) and of the detailed radiative transfer (taking into account all significant sources of opacity and emission and thereby providing a proper treatment of line blocking[2] and blanketing[3]) as well as of the temperature structure (determined via a solution of the microscopic energy equation which states that the energy – taking into account the local energy production rates from the radioactive decay chains – must be conserved in the atmosphere (cf. Pauldrach et al. 2013)). As part of the selfconsistent solution of the entire system one obtains the frequency-resolved radiation field at the outer boundary – the synthetic emergent spectrum – which can, in principle, be directly compared to observations.

Although such spectral synthesis codes are being applied with great success to the analysis of spectra of hot stars with winds, reproducing the stellar spectra in great detail, the modelling of SN Ia spectra has not yet reached the same level of realism. It is the complex physical conditions within the expanding SN Ia ejecta compared to those in the atmospheres of hot stars which make it extremely difficult to implement models that allow reliable and quantitative analysis. Despite the underlying microphysics being the same in stellar atmospheres and supernova ejecta, the physical conditions are sufficiently different that approximations which are excellent in stellar atmospheres become invalid in supernovae.

One such is the commonly applied diffusion approximation as the inner boundary in the radiative transfer modelling, which breaks down because the opacity in SN Ia is dominated by electron scattering over large wavelength ranges, in contrast to stars which have a strong free-free and bound-free continuum. Electron scattering, however, does not couple the radiation field to the thermal pool, so that the radiation field cannot become Planckian even at depth. We have developed a semi-analytical description that allows us to overcome some of the limiting assumptions in the conventional treatment of the lower boundary in SN Ia radiative transfer models. The improved boundary condition explicitly takes into account the dominance of scattering processes at the longer wavelengths and leads to a much better agreement with observed spectra in the near-infrared part (cf. Sauer et al. 2006).

Still, the models tend to predict much more flux in the red and near-infrared part of the spectrum than is actually observed, and this behavior must be connected to the radiative transfer formalism which is effectively thwarted by the superposition of thousands of strong metal absorption lines blocking the flux in the UV part of the spectrum within the expanding envelope (cf. Sect. 5). We have identified two severe problems in the radiative transfer related to these extreme conditions. The first one is connected to a hidden form of lambda iteration which is caused by the mutual interaction of strong spectral lines and which is inherently connected to the line blocking effect. Based on this finding we have deduced a solution to this problem, such that the formation of these lines results in a correct description of the pseudo-continuum. With the application of this method the changes obtained in the energy distributions and line spectra lead to a much better agreement with the observed spectra than those of previous models (cf. Pauldrach et al. 2013).

---

[1] Involving different levels of complexity, models have been developed by Branch et al. (1985); Mazzali et al. (1993); Mazzali & Lucy (1993); Eastman & Pinto (1993); Höflich et al. (1995); Nugent et al. (1995); Pauldrach et al. (1996); Nugent et al. (1997); Lentz et al. (2001); Höflich (2005); Stehle et al. (2005); Sauer et al. (2006); Baron et al. (2006); Kasen et al. (2006). With regard to the purpose of the specific codes various simplifications have been applied, as not all approaches are intended to provide a comprehensive description of the time-dependent spectra, including the detailed statistical equilibrium of all relevant elements.

[2] The effect of line blocking refers to an attenuation of the radiative flux in the EUV and UV due to the combined opacity of a huge number of metal lines present in SN Ia in these frequency ranges. It drastically influences ionization and excitation rates due to its impact on the radiation field through radiative absorption and scattering processes.

[3] As a consequence of line blocking, only a small fraction of the radiation is re-emitted and scattered in the outward direction. Most of the energy is radiated back to deeper layers of the SN Ia, leading there to an increase of the temperature ("backwarming"). Due to this increase of the temperature, more radiation is emitted at lower energies, an effect known as line blanketing.

The second problem we have encountered in our investigation of the behavior of the radiative transfer under the extreme conditions of SN Ia envelopes is related to the standard Feautrier solution, which, in the form it is used, fails to correctly describe the radiation field in the presence of large radial opacity jumps.

### 2.1. The Feautrier solution of the radiative transfer equation

For the solution of the radiative transfer in the observer's frame, three similar algorithms are used in our procedure. The start solution is obtained from a Rybicki-method (Rybicki 1971; Mihalas 1978, p. 158) that implicitly solves the transport equation including Thomson scattering. Because this solution does not have a very high accuracy, an iteration for the radiation field that enters into the Thomson emissivity is performed. Here, the ray-by-ray solution and the angle-integrated moment equations are iterated. Both systems are solved with a Feautrier scheme (cf. Mihalas 1978, p. 156) and the iteration is performed twice for each frequency point: first for a pure continuum model and afterwards for the full problem with continuum and spectral lines. One crucial limitation of this approach is that, strictly speaking, the Feautrier method is only applicable if the opacities and source functions vary only slowly over radius, i.e., the radial variation of the lines would have to be resolved by the grid. The sampling iteration, however, uses a coarse radial grid consisting of a limited number of depth points. Therefore, most lines will be present at a single grid point only (being Doppler-shifted to different frequency points at the neighboring depth points) and the standard procedure thus has to be modified to be able to use this algorithm (see also Section 2.3). In order to understand how the method fails and to determine what modifications are required the procedure however has to be presented in some detail first.

For the numerical description of the transfer equation a discretization method which admits a sufficient accuracy must be applied and the solution is then carried out in the usual Cartesian $p$-$z$-coordinate system (Fig. 1), in which each $p$-ray corresponds to a $\mu$-direction in spherical coordinates. The transformation between the $r$-$\mu$ and $p$-$z$-coordinates is

$$\mu = \frac{z}{r} \quad \Rightarrow \quad \mu = \frac{z}{\sqrt{z^2 + p^2}} \tag{1}$$

$$r^2 = z^2 + p^2 \quad \Rightarrow \quad r = \sqrt{p^2 + z^2} \tag{2}$$

and, therefore,

$$dr = \frac{z}{\sqrt{z^2 + p^2}} dz = \mu \, dz \tag{3}$$

$$d\mu = \left(\frac{1}{\sqrt{p^2 + z^2}} - \frac{z^2}{\sqrt{p^2 + z^2}^3}\right) dz = \frac{1 - \mu^2}{r} dz. \tag{4}$$

To solve the transport equation in this $p$-$z$ geometry it is useful to convert the radiative transfer equation $dI/dz = -\chi I + \eta$, a first-order differential equation with a single boundary condition, into a second-order differential equation with a boundary condition for each side (e.g., Mihalas 1978). For this method, introduced by P. Feautrier, the transfer equation for each $p$-ray is rewritten for the intensities in positive and negative $z$-direction

$$\frac{dI^\pm}{d\tau} = \pm(S - I^\pm) \quad \text{with} \quad d\tau = -\chi dz. \tag{5}$$

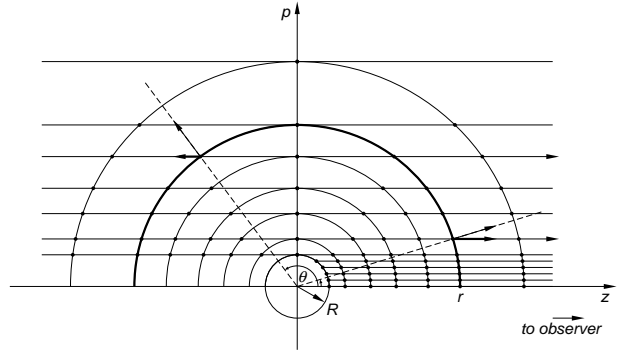

**Fig. 1.** Sketch of the $p$-$z$ coordinate system that is used for the solution of the radiative transfer. In this geometry, each $p$-ray corresponds to a ray in direction $\mu$ in the $r$-$\mu$ coordinate system.

By introducing the new variables

$$\begin{aligned} u &= \tfrac{1}{2}(I^+ + I^-) \quad \text{(intensity-like)} \\ v &= \tfrac{1}{2}(I^+ - I^-) \quad \text{(flux-like)} \end{aligned} \tag{6}$$

Eq. 5 can be written as a system of two coupled differential equations,

$$\frac{du}{d\tau} = v \quad \text{and} \quad \frac{dv}{d\tau} = u - S, \tag{7}$$

and after combining these two equations by eliminating $v$, one obtains the single second-order differential equation for $u$ with the independent variable $\tau$:

$$\frac{d^2 u}{d\tau^2} = u - S. \tag{8}$$

In the standard discretized scheme, this equation is represented as a set of difference equations – one for each radial grid point $i$ along the ray:

$$\left.\frac{d^2 u}{d\tau^2}\right|_{\tau_i} \approx \frac{\left.\frac{du}{d\tau}\right|_{\tau_{i+1/2}} - \left.\frac{du}{d\tau}\right|_{\tau_{i-1/2}}}{\tau_{i+1/2} - \tau_{i-1/2}}$$

$$\approx \frac{\frac{u_{i+1} - u_i}{\tau_{i+1} - \tau_i} - \frac{u_i - u_{i-1}}{\tau_i - \tau_{i-1}}}{\tfrac{1}{2}(\tau_{i+1} - \tau_i) - \tfrac{1}{2}(\tau_i - \tau_{i-1})}. \tag{9}$$

This represents a linear equation system for $u_i$

$$A_i u_{i-1} + B_i u_i + C_i u_{i+1} = S_i \tag{10}$$

with the coefficients

$$\begin{aligned} A_i &= -\left(\tfrac{1}{2}(\tau_{i+1} - \tau_{i-1})(\tau_i - \tau_{i-1})\right)^{-1} \\ C_i &= -\left(\tfrac{1}{2}(\tau_{i+1} - \tau_{i-1})(\tau_{i+1} - \tau_i)\right)^{-1} \\ B_i &= 1 - A_i - C_i. \end{aligned} \tag{11}$$

This tridiagonal matrix system can be solved efficiently by standard linear algebra solvers. In our non-LTE-models a Rybicki-type scheme (Rybicki 1971; Mihalas 1978, p. 158) is used that solves all $p$-rays simultaneously to implicitly account for the radiation field $J$ that enters in the scattering part of the source function $S$ (cf. Appendix A of Pauldrach et al. 2001).

As equation Eq. 9 contains only differences in $\tau$ that can be easily derived from the opacities $\chi_i$ at the respective grid points of the $z$ grid from

$$\Delta\tau_{i,i-1} = -\tfrac{1}{2}(\chi_i + \chi_{i-1})(z_i - z_{i-1}), \quad (12)$$

the system is well behaved if opacities and source functions are slowly varying functions of $z$. Problems can, however, arise if these conditions cannot be guaranteed, such as when strong ionization edges occur or a large velocity gradient shifts strong lines in frequency. Such situations can cause large variations of the opacity from grid point to grid point at a given frequency. One such problematic situation[4] – already discussed in Pauldrach et al. (2001) – occurs if a point with a large source function $S_i$ and low opacity $\chi_i$ is adjacent to a point with high opacity $\chi_{i+1}$ and low or average source function $S_{i+1}$. In reality, the large source function $S_i$ should have only a small impact on the radiation field as it occurs in a low opacity region (therefore the emissivity is also small). In the solution of the equation system, however, the emission computed in this situation is on the order of

$$\Delta I \approx \bar{S}\Delta\tau \approx \tfrac{1}{2}(S_{i+1} + S_i)\cdot\tfrac{1}{2}(\chi_{i+1} + \chi_i)(z_{i+1} - z_i). \quad (13)$$

In this expression, the term $S_i\chi_{i+1}$ dominates under the assumed conditions, which leads to an artificially enhanced emission. This happens because physically it does not make sense to multiply the source function of one point with the opacity of another point. Source functions, as a matter of principle, are only meaningful quantities relative to the opacity at the same point. As discussed by Pauldrach et al. (2001), the method can still be used if the equation system is written in $z$ instead of $\tau$ (as in Feautrier's original paper) because the $z$-dependence of $\chi$ can be treated correctly only in this formulation. Eq. 8 then becomes

$$\frac{1}{\chi}\frac{d}{dz}\left(\frac{1}{\chi}\frac{du}{dz}\right) = u - S. \quad (14)$$

Carrying out the same discretization as before leads to the new difference system

$$\frac{1}{\chi_i}\left(\frac{d}{dz}\left(\frac{1}{\chi}\frac{du}{dz}\right)\right)\bigg|_i \approx \frac{1}{\chi_i}\frac{\left(\frac{1}{\chi}\frac{du}{dz}\right)_{i+1/2} - \left(\frac{1}{\chi}\frac{du}{dz}\right)_{i-1/2}}{z_{i+1/2} - z_{i-1/2}}$$

$$\approx \frac{1}{\chi_i}\frac{\frac{1}{\bar{\chi}_{i+1,i}}\frac{u_{i+1}-u_i}{z_{i+1}-z_i} - \frac{1}{\bar{\chi}_{i,i-1}}\frac{u_i-u_{i-1}}{z_i-z_{i-1}}}{\tfrac{1}{2}(z_{i+1}+z_i) - \tfrac{1}{2}(z_i+z_{i-1})}. \quad (15)$$

Thus, the coefficients Eq. 11 are now

$$A_i = -\left(\tfrac{1}{2}\chi_i(z_{i+1}-z_{i-1})\bar{\chi}_{i,i-1}(z_i-z_{i-1})\right)^{-1}$$
$$C_i = -\left(\tfrac{1}{2}\chi_i(z_{i+1}-z_{i-1})\bar{\chi}_{i+1,i}(z_{i+1}-z_i)\right)^{-1}$$
$$B_i = 1 - A_i - C_i, \quad (16)$$

where the the first terms of $A$ and $B$ now contain the *local* opacity. For the mean opacity $\bar{\chi}$, the geometric mean

$$\bar{\chi}_{i+1,i} = \sqrt{\chi_{i+1}\cdot\chi_i} \quad (17)$$

generally produces reasonable results for stars. While Eq. 17 puts more weight on the lower of the two opacities, using the arithmetic mean

$$\bar{\chi}_{i+1,i} = \tfrac{1}{2}(\chi_i + \chi_{i+1}) \quad (18)$$

weights the opacities equally. Another caveat about using the Feautrier scheme for solving the radiative transport for conditions where the opacity is not smoothly distributed over radius is that part of the line emissions are incorrectly re-absorbed in the adjacent intervals because the source functions are defined on the grid points only. For a detailed discussion of this issue, see Section 2.3.

**Boundary conditions** To close the system in Eq. 9 appropriate boundary conditions have to be chosen. The respective boundary equations can be obtained from an expansion of $u_i$ in $\tau$ at the boundary points outside ($i = N$) and inside ($i = 1$):

$$u_{N-1} = u_N + (\tau_N - \tau_{N-1})\frac{du}{d\tau}\bigg|_{\tau_N} + \tfrac{1}{2}(\tau_N - \tau_{N-1})^2\frac{d^2u}{d\tau^2}\bigg|_{\tau_N} \quad (19)$$

and

$$u_2 = u_1 + (\tau_2 - \tau_1)\frac{du}{d\tau}\bigg|_{\tau_1} + \tfrac{1}{2}(\tau_2 - \tau_1)^2\frac{d^2u}{d\tau^2}\bigg|_{\tau_1}. \quad (20)$$

It is assumed that there is no incident radiation ($I^- \equiv 0$) at the outer boundary[5]. Thus from Eq. 7, one obtains

$$u = v \quad \Rightarrow \quad \frac{du}{d\tau}\bigg|_{\tau=0} = u. \quad (21)$$

With Eq. 8, this leads to the outer boundary condition

$$u_{N-1} = u_N + (\tau_{N-1} - \tau_N)u_N + \tfrac{1}{2}(\tau_{N-1} - \tau_N)^2(u_N - S_N) \quad (22)$$

and the coefficients

$$A_N = -2(\tau_{N-1} - \tau_{N-1})^{-2} \quad (23)$$
$$B_N = 1 + 2(\tau_{N-1} - \tau_N)^{-1} + 2(\tau_{N-1} - \tau_N)^{-2}$$
$$= 1 + 2(\tau_2 - \tau_1)^{-1} - A_N. \quad (24)$$

At the inner boundary, rays that intersect the core ($p < R$) must be distinguished from those that do not ($p > R$). For core-rays the incident intensity has to be explicitly specified $I^+ = I_{\text{core}}$, while for non-core rays a reflecting boundary $I^+ = I^-$ is used. Noting that $v = \tfrac{1}{2}(I^+ - I^- + I^+ - I^+) = I^+ - u$, one gets in Eq. 7

$$\frac{du}{d\tau}\bigg|_{\tau_{\max}} = I_{\text{core}} - u \quad (p < R) \quad (25)$$

for core-rays and

$$\frac{du}{d\tau}\bigg|_{\tau_{\max}} = 0 \quad (p > R) \quad (26)$$

for non-core rays. Combining Eq. 19, Eq. 8 and, Eq. 7 one gets

$$u_2 = u_1 + (\tau_1 - \tau_2)(I_{\text{core}} - u_1)$$
$$\quad + \tfrac{1}{2}(\tau_2 - \tau_1)^2(u_1 - S_2) \quad (p < R) \quad (27)$$
$$u_2 = u_1 + \tfrac{1}{2}(\tau_2 - \tau_1)^2(u_1 - S_1) \quad (p > R) \quad (28)$$

---

[4] We note that Feautrier presented his method in the form of eq. 14, which does not suffer from this first problem. It is the description using the mixed "$\tau$"s (such as given by Mihalas) which leads to these errors.

[5] For stellar atmosphere models, at the outer boundary an extrapolation for $I^-$ from radii larger than the computational grid is performed. This is not relevant for supernovae because of steep density gradients and low absolute densities in the outer region.

and the coefficients

$$C_1 = -2(\tau_1 - \tau_2)^{-2} \tag{29}$$
$$B_1 = 1 - 2(\tau_1 - \tau_2)^{-1} + 2(\tau_1 - \tau_2)^{-2} \quad (p < R) \tag{30}$$
$$B_1 = 1 + 2(\tau_1 - \tau_2)^{-2} \quad (p > R) \tag{31}$$
$$S_1^* = S_1 - 2(\tau_1 - \tau_2)^{-1} I_{\text{core}} \quad (p < R) \tag{32}$$
$$S_1^* = S_1 \quad (p > R). \tag{33}$$

These boundary conditions can be expressed in terms of the $z$-variable analogous to the non-boundary equations. For the issues to be discussed in the present paper, it is sufficient, at least for the test cases, to assume that the incident intensity $I_{\text{core}}$ is given. (Note that the much more complex incident intensity which actually is used in the non-LTE models (cf. Sect. 5) is discussed in detail by Sauer et al. 2006.)

## 2.2. Solution of the moment equations

A similar system can be employed to solve the moment equations

$$\frac{d\tilde{H}}{d\tilde{\tau}} = \frac{1}{q}(\tilde{J} - \tilde{S}), \quad \frac{d(qf\tilde{J})}{d\tilde{\tau}} = \tilde{H} \tag{34}$$

$$\frac{d^2(qf\tilde{J})}{d\tilde{\tau}^2} = \frac{1}{q}(\tilde{J} - \tilde{S}) \tag{35}$$

where all symbols have their usual meanings. (Again the index $\nu$ has been dropped for brevity.) The last equation, Eq. 35, can be written in terms of the radius $r$

$$\frac{1}{q\chi}\frac{d}{dr}\left(\frac{1}{q\chi}\frac{d(qf\tilde{J})}{dr}\right) = \frac{1}{q}(\tilde{J} - \tilde{S}). \tag{36}$$

The additional advantage here is that one can implicitly solve for the contribution of Thomson scattering to the source function $S$. Separating the emissions due to true processes and those due to Thomson-scattering ($\eta^{\text{Th}} = \chi^{\text{Th}} J$), one can write

$$S = \frac{\eta^{\text{true}}}{\chi^{\text{true}} + \chi^{\text{Thoms}}} + \frac{\chi^{\text{Thoms}}}{\chi^{\text{true}} + \chi^{\text{Thoms}}} J = S^\dagger + \beta J \tag{37}$$

with the definitions

$$S^\dagger := \frac{\eta^{\text{true}}}{\chi} \quad \text{and} \quad \beta := \frac{\chi^{\text{Thoms}}}{\chi}. \tag{38}$$

Using this in Eq. 36 gives

$$\frac{1}{\chi}\frac{d}{dr}\left(\frac{1}{q\chi}\frac{d(qf\tilde{J})}{dr}\right) = \tilde{J}(1-\beta) - \tilde{S}^\dagger \tag{39}$$

which can be discretized by

$$-\frac{1}{\chi_i(r_{i+1}-r_{i-1})}\left(\frac{f_{i+1}q_{i+1}\tilde{J}_{i+1} - f_i q_i \tilde{J}_i}{\overline{q\chi}_{i+1,i}(r_{i+1}-r_i)} - \frac{f_i q_i \tilde{J}_i - f_{i-1}q_{i-1}\tilde{J}_{i-1}}{\overline{q\chi}_{i,i-1}(r_i - r_{i-1})}\right)$$
$$+ (1-\beta)\tilde{J}_i = S_i^\dagger, \tag{40}$$

where $\overline{q\chi}_{i+1,i}$ are again appropriate means of the product $q\chi$ at adjacent grid points. However, the system still needs the Eddington factors $f_\nu = K_\nu/J_\nu$ as an external input obtained from the ray-by-ray solution and the coefficients of the equation system

$$\mathcal{A}_i \tilde{J}_{i+1} + \mathcal{B}_i \tilde{J}_i + C_i \tilde{J}_{i-1} = \mathcal{K}_i \tag{41}$$

become

$$\mathcal{A}_i = -\frac{f_{i+1}q_{i+1}}{\frac{1}{2}\chi_i(r_{i+1}-r_{i-1})\overline{q\chi}_{i+1,i}(r_{i+1}-r_i)} \tag{42}$$

$$C_i = -\frac{f_{i-1}q_{i-1}}{\frac{1}{2}\chi_i(r_{i+1}-r_{i-1})\overline{q\chi}_{i,i-1}(r_i - r_{i-1})} \tag{43}$$

$$\mathcal{B}_i = \frac{f_i q_i}{\frac{1}{2}\chi_i(r_{i+1}-r_{i-1})}\left(\frac{1}{\overline{q\chi}_{i+1,i}(r_{i+1}-r_i)} - \frac{1}{\overline{q\chi}_{i,i-1}(r_i-r_{i-1})}\right)$$
$$+ (1-\beta) \tag{44}$$

$$\mathcal{K}_i = \tilde{S}^\dagger. \tag{45}$$

At the boundaries, the system is closed by employing factors similar to the second Eddington factor[6]

$$h := \frac{\int_0^1 u(\mu)\mu \, d\mu}{\int_0^1 u(\mu) \, d\mu} \tag{46}$$

with $u(\mu)$ from the solution of the ray-by-ray solution. At the outer boundary, since $u(\tau = 0) \equiv v(\tau = 0)$, this is just

$$h(\tau = 0) = \left.\frac{H}{J}\right|_{\tau=0}. \tag{47}$$

Thus, the outer boundary equation is

$$\left.\frac{d(fq\tilde{J})}{d\tilde{\tau}}\right|_{\tau=0} = h(\tau = 0)\tilde{J}(\tau = 0). \tag{48}$$

The inner boundary ($r = R$) is treated similarly; however, because $\int u\mu \, d\mu \neq H$, $I_{\text{core}}$ from the ray-by-ray solution has to be employed here as well, noting that

$$H(\tau_{\max}) = \int_0^1 v\mu \, d\mu = \int_0^1 I_{\text{core}}\mu \, d\mu - \int_0^1 u\mu \, d\mu. \tag{49}$$

This results in

$$\left.\frac{d(fq\tilde{J})}{d\tilde{\tau}}\right|_{\tau_{\max}} = R^2 \int_0^1 I_{\text{core}}\mu \, d\mu - h(\tau_{\max})\tilde{J}. \tag{50}$$

The coefficients at the outer boundary are

$$C_N = -\frac{f_{N-1}q_{N-1}}{\overline{q\chi}_{N,N-1}(r_{N-1}-r_N)} \tag{51}$$

$$\mathcal{B}_N = \frac{f_N q_N}{\overline{q\chi}_{N,N-1}(r_{N-1}-r_N)} + h_N \tag{52}$$

$$\mathcal{K}_N = \tilde{S}^\dagger \tag{53}$$

and at the inner boundary are

$$\mathcal{A}_1 = -\frac{f_2 q_2}{\overline{q\chi}_{1,2}(r_2 - r_1)} \tag{54}$$

$$\mathcal{B}_1 = \frac{f_1 q_1}{\overline{q\chi}_{1,2}(r_2 - r_1)} + h_1 \tag{55}$$

$$\mathcal{K}_1 = \int_0^1 I_{\text{core}}\mu \, d\mu. \tag{56}$$

Again, we will assume $I_{\text{core}}$ to be given, and we refer to Sauer et al. (2006) for the details of the inner boundary intensity actually used in the models. The solution of this tridiagonal matrix scheme is again performed by efficient BLAS-functions.

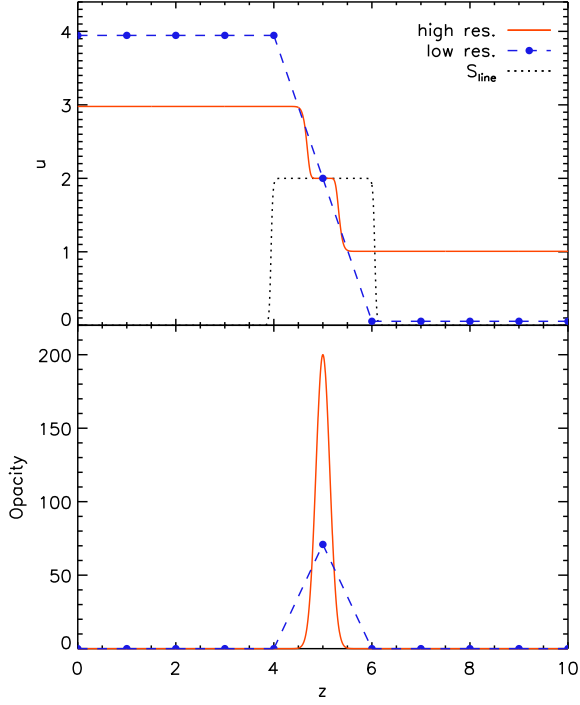

**Fig. 2.** Toy model to illustrate the problem of self absorption in the standard Feautrier scheme for a single *p*-ray (see text). The lower panel shows the opacity distribution for this model. The opacities of the models are chosen such that the same total optical depth over the entire radial interaction zone is reached. The upper panel shows the *u* function derived from the standard Feautrier solution using an incident intensity $I^+ = 4$ from the left side and $I^- = 0$ on the right side. The black lines denote the run of the $S_{\text{line}}$.

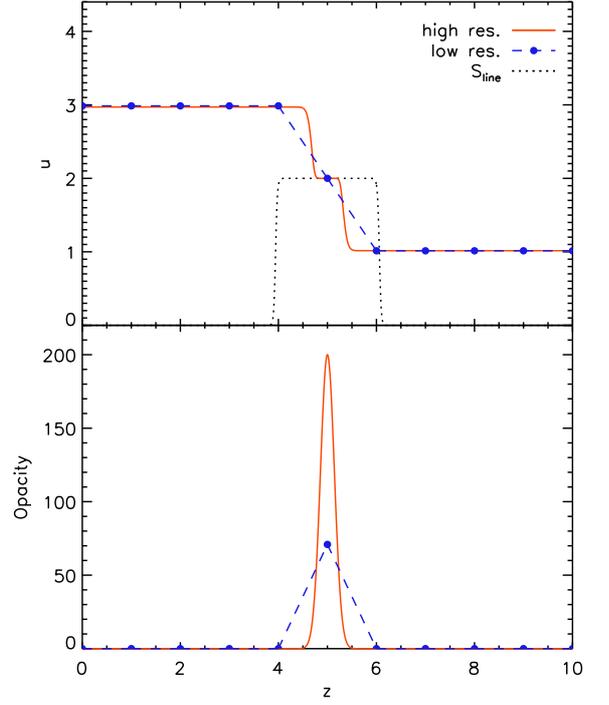

**Fig. 3.** Single-ray models corresponding to Fig. 2 with the $\xi$-correction applied. While the high-resolution model remains unaffected, the model on the coarse grid now reproduces the analytically expected result much better.

### 2.3. Self-absorption of lines in the Feautrier scheme

As mentioned in Section 2, the Feautrier scheme for the solution of the radiative transfer leads to difficulties if used for physical conditions where the opacities and emissivities have strong variations at adjacent depth points, such as in the envelopes of SN Ia.

In the Feautrier scheme, the total source functions are evaluated on the respective radial grid points. This effectively sets the entire emission of an interval on that grid-point, while the transport coefficients contain the absorption over an interval between grid points. Consider the situation where one grid point $i$ has a low opacity $\chi_i$ (which makes the source function $S_i$ at this point meaningless), and further assume that the adjacent point $i-1$ has a large opacity $\chi_{i-1}$ and a source function $S_{i-1}$ (e.g., by the presence of a strong line). In reality the emission from a line occurs at a distance in $r$ where the line becomes optically thick $\tau > 1$, which means that the low opacity point would "see" the source function of the line at the point where $\Delta\tau \gtrsim 1$ is reached. The numerical solution, however, systematically underestimates the radiation being transported from the point $i-1$ into the low opacity region $i$. This effect can be understood as follows: recall the general structure of the equation system solved (cf. Section 2):

$$a_i u_{i-1} + b_i u_i + c_i u_{i+1} = q_i S_i = k_i \tag{57}$$

where the coefficients are such that

$$b_i = q_i - a_i - c_i. \tag{58}$$

---
[6] The second Eddington factor is actually defined as the ratio $(H/J)|_{\tau=0}$ at the outer boundary of the atmosphere.

The coefficients $a_i$ and $c_i$ relate to the transport over the interval $[i, i-1]$ and $[i, i+1]$ respectively, while $b_i$ and $q_i$ describe the local grid point $i$. Thus the former are functions of the "remote" optical depth interval $\Delta\tau_i^\pm = \bar{\chi}_{i,i\pm 1}\Delta z_{i,i\pm 1}$, while $q_i = \Delta\tau_i$ is a function of the "local" optical depth interval $\Delta\tau_i = \chi_i \Delta z_{i+1,i-1}$:

$$a_i = (\Delta\tau_i^-)^{-1}, \qquad c_i = (\Delta\tau_i^+)^{-1}, \qquad b_i = q_i - a_i - c_i. \tag{59}$$

Let us now consider the extreme case of a single line in a low opacity environment by setting $S_{i-1} \neq 0$, $\chi_{i-1} \neq 0$, $S_{j\leq i} \to 0$, and $\chi_{j\leq i} \to 0$. For the coefficients in Eq. 57 this means that

$$\begin{aligned}
\Delta\tau_i^- \gg 1 &\Rightarrow a_i \to 0 \\
\Delta\tau_i^- \ll 1 &\Rightarrow c_i \gg 1 \\
\Delta\tau_i \to 0 &\Rightarrow q_i \to 0 \\
&\Rightarrow b_i = -c_i.
\end{aligned} \tag{60}$$

Therefore, the equation for $u_i$ is

$$u_i = -\frac{c_i}{b_i} u_{i+1} = u_{i+1}. \tag{61}$$

All points $j > i-1$ decouple entirely from the point $i-1$ and accordingly, $u_j$ is determined by whatever is emitted. Fig. 2 shows this example; from the left, an incident "core" intensity is assumed, whereas the right boundary has no incoming radiation.

For this example, two models have been computed: a high resolution model with a small grid spacing that resolves the actual line profile, and which we can reasonably assume to yield results very close to the exact solution since it avoids the problems described above; and a second low-resolution model that corresponds more to the conditions present in the radiative transfer solver used for the sampling method (cf. Section 2). The high-resolution model assumes a Gaussian profile for the line, which in the low-resolution model is represented by a single grid point. The opacity at that point is chosen such that the optical depth through the radial interaction zone is the same as for the highly resolved model. The opacity distribution of both models on the ray are shown in the lower panel. The upper panel shows the $u$-function derived from the Feautrier algorithm described in Section 2 and the adopted source function $S_{\text{line}} = 2$ in the line for both models (black dashed line). For the left side, an incident intensity $I^+ = 4$ was assumed, while the right side had $I^- = 0$. The red line denotes the solution of the highly resolved model and follows what one would expect for that case theoretically: within the optically thick line $u$ approaches $S_{\text{line}}$, while on the right side it reaches $\frac{1}{2} S_{\text{line}}$ because $u = \frac{1}{2}(I^+ + I^-) = \frac{1}{2}(S_{\text{line}} \Delta \tau + 0)$. Accordingly, on the side where an incident $I^+$ is present, the result is half way between $I^+$ and $S_{\text{line}}$ ($u = \frac{1}{2}(I^+ + S_{\text{line}} \Delta \tau)$). The blue line with the bullet points shows the solution for the model on the coarse grid. (The values are only derived on the grid points as indicated; the connecting lines are just for better visibility.) As can be clearly seen, on both sides of the line the solution significantly underestimates the radiation field compared to the exact solution. Effectively, the line emission is not represented in the transport at all because the solutions outside the line are only coupled to the boundary points. Of course an extreme case like the one presented here will rarely occur in real models because the continuous opacity will usually smooth out the profiles and also prevent zero opacity at some points. However, in particular for supernova where the line opacity dominates the total opacity over almost the entire spectrum, cases similar to this are more likely to occur. Here the standard solver will systematically underestimate the radiation field in the (radial) gaps between lines and toward the outer boundary, leading to a loss of total radiative energy that is not represented in the rate equations. Note, that this only affects the transport in our method I for the radiative transport. The detailed solution is derived on a micro-grid that resolves the line profiles in small $\Delta\tau$-steps and therefore does *not* suffer from this problem (cf. Pauldrach et al. 2001).

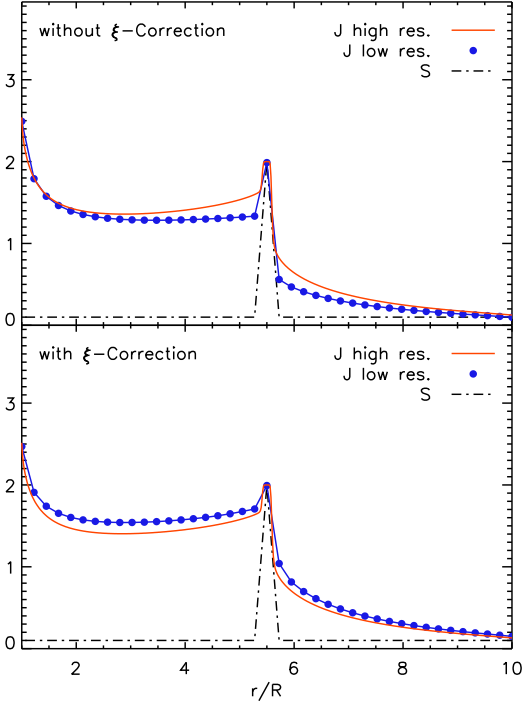

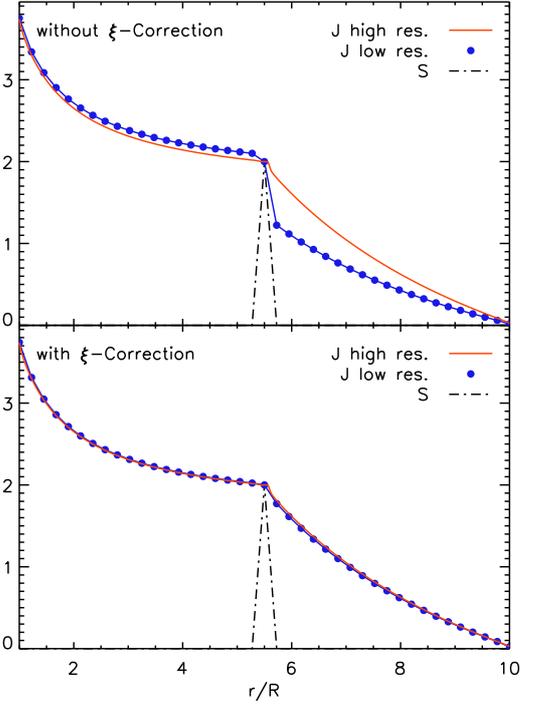

**Fig. 4.** Spherically symmetric test models for the corrected Feautrier coefficients. In all models the incident intensity at the core was set to $I^+ = 4$, while at the outer boundary no incoming radiation was assumed. The upper panel shows the uncorrected model and the lower panel shows the same setup with the correction applied. In this model we assumed an optically thick line ($\tau_{\text{line}} = 50$), and a low background true opacity and source function.

**Fig. 5.** Same setup as in 4, but instead with Thomson opacity and emissivity.

## 3. An improved solution method for the Feautrier scheme

As a primary constraint the correction method should not affect the transport of radiation through a grid point. Additionally, the solution has to remain unchanged for a smooth, continuous run of opacities because these cases are correctly represented by the standard scheme. Additionally, at high grid resolution, both the new and the old scheme should agree on the same solution because in the limit of infinitely small intervals the standard method approaches the exact solution. We found that the first constraint strictly excludes any modification of the transport terms $a_i$ and $c_i$. Therefore, the only remaining option is to adjust the source terms and the $b$-coefficient. The concept is to correct the intensity derived at a particular grid point by adding an additional contribution from the emission of adjacent grid points. The amount of correction has to be a function of the local $\Delta\tau_i$ and the respective remote $\Delta\tau_i^\pm$. Thus, the following approach for a new coefficient $b_i$ and the right-hand side $k_i$ of Eq. 57 has been made:

$$k_i = q_i S_i + \xi_i^+ S_{i+1}^* + \xi_i^- S_{i-1}^* \tag{62}$$
$$b_i = q_i - a_i - c_i + \xi_i^+ + \xi_i^-, \tag{63}$$

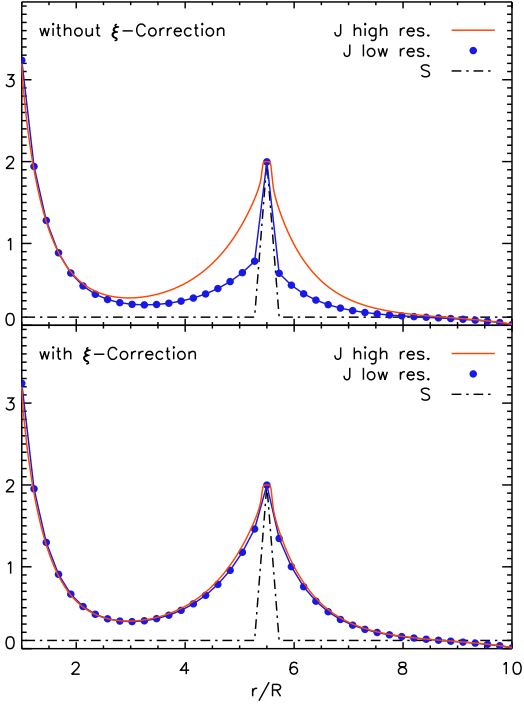

**Fig. 6.** Same setup as in 4, but including both true opacity and source function as well as Thomson opacity and emissivity as background.

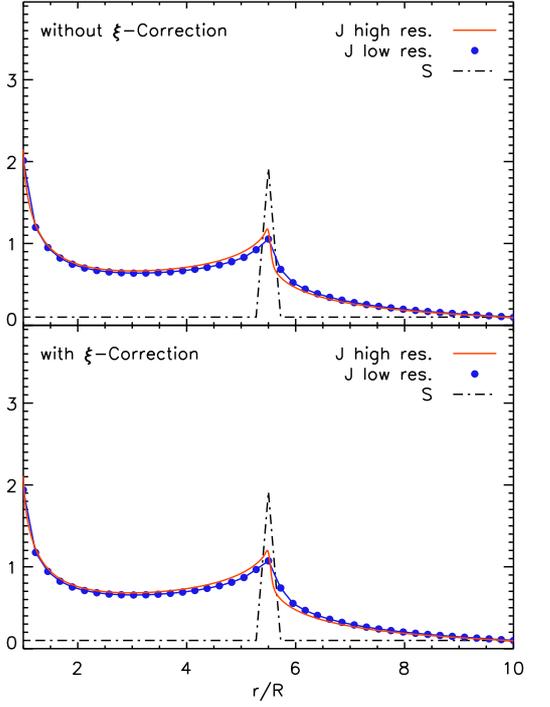

**Fig. 7.** Same situation as in Fig. 6, but for an optically much thinner line ($\tau_{\text{line}} = 0.5$). Here the old and new method as expected produce very similar results.

with correction functions $\xi_i^\pm = \xi_i^\pm(\Delta\tau_i, \Delta\tau_i^\pm)$ to be determined. The source functions $S^*$ are taken to be the (total) line source functions $\eta_{\text{line}}/\chi_{\text{line}}$ at the respective point to ensure that the correction does not affect the continuum transport. To obtain the original coefficients of Eq. 10 in Section 2, the system Eq. 59 has to be divided by $q_i$. Thus, coefficients $A_i$ and $C_i$ in Eq. 16 remain unchanged, while $B_i$ and the right-hand side $K_i$ acquire additional terms:

$$A_i = -\left(\tfrac{1}{2}\chi_i(z_{i+1} - z_{i-1})\bar\chi_{i,i-1}(z_i - z_{i-1})\right)^{-1} \tag{64}$$

$$C_i = -\left(\tfrac{1}{2}\chi_i(z_{i+1} - z_{i-1})\bar\chi_{i+1,i}(z_{i+1} - z_i)\right)^{-1} \tag{65}$$

$$B_i = 1 - A_i - C_i + \left(\tfrac{1}{2}\chi_i(z_{i+1} - z_{i-1})\right)^{-1}(\xi_i^+ + \xi_i^-) \tag{66}$$

$$K_i = S_i + \left(\tfrac{1}{2}\chi_i(z_{i+1} - z_{i-1})\right)^{-1}\left(\xi_i^+ S_{i+1}^* + \xi_i^- S_{i-1}^*\right). \tag{67}$$

The coefficients at the boundary conditions have to be adjusted accordingly using the corresponding adjacent interval.

### 3.1. Moment equations

In the solution scheme for the moment equations Eq. 41 and Eq. 42, a similar method has to be applied. For consistency with the ray-by-ray solution, the correction terms $\xi^\pm u_i$ are integrated over $\mu\,d\mu$ introducing a factor $h_i' = \int u_i(\mu)\mu\,d\mu/J_i$ analogous to $h$ in Eq. 47. Here, however, the numerical integration for $h_i'$ has to be performed using integration weights on the respective grid point instead of the weights on the intermesh points used for the flux integration. Eventually, one derives the new coefficients for the system Eq. 41 to be

$$\mathcal{A}_i = -\frac{f_{i+1}q_{i+1}}{\tfrac{1}{2}\chi_i(r_{i+1} - r_{i-1})\overline{q\chi}_{i+1,i}(r_{i+1} - r_i)} \tag{68}$$

$$\mathcal{C}_i = -\frac{f_{i-1}q_{i-1}}{\tfrac{1}{2}\chi_i(r_{i+1} - r_{i-1})\overline{q\chi}_{i,i-1}(r_i - r_{i-1})} \tag{69}$$

$$\mathcal{B}_i = \frac{f_i q_i}{\tfrac{1}{2}\chi_i(r_{i+1} - r_{i-1})}\left(\frac{1}{\overline{q\chi}_{i+1,i}(r_{i+1} - r_i)} - \frac{1}{\overline{q\chi}_{i,i-1}(r_i - r_{i-1})}\right)$$
$$+ (1 - \beta) + \left(\tfrac{1}{2}\chi_i(r_{i+1} - r_{i-1})\right)^{-1}(\xi_i^+ + \xi_i^-)h_i'r_i^2 \tag{70}$$

$$\mathcal{K}_i = \tilde{S}^\dagger + \tfrac{1}{2}\left(\xi_i^+ S_{i+1}^* + \xi_i^- S_{i-1}^*\right)\left(\tfrac{1}{2}\chi_i(r_{i+1} - r_{i-1})\right)^{-1}r_i^2. \tag{71}$$

The first-order boundary conditions for the system remain unchanged because no source functions enter here.

### 3.2. Correction functions $\xi^\pm$

Now that we know where the correction is to be applied, suitable functions $\xi^\pm$ that depend on the opacities in the "local" and "remote" intervals have to be found. Correction is only needed for the case where the respective remote and the local $\tau$-interval are low. In this case, if the line source function $S^*$ of the adjacent point is large, the emission of the local point has to be enhanced. In addition, the correction function has to drop to zero for large "remote" $\Delta\tau$ faster than the $1/\Delta\tau$-terms in the transport coefficients.

A possible choice for the $\xi$-functions that has been found to provide the required properties is

$$\xi_i^- = \left(1 - e^{-\bar\chi_{\text{line}}^-(z_i - z_{i-1})}\right)^2 e^{-\Delta\tau_i}$$
$$\xi_i^+ = \left(1 - e^{-\bar\chi_{\text{line}}^+(z_{i+1} - z_i)}\right)^2 e^{-\Delta\tau_i} \tag{72}$$

with the averaged opacities

$$\bar\chi_{\text{line}}^\pm = \tfrac{1}{2}\left(\chi_i^{\text{line}} + \chi_{i\pm1}^{\text{line}}\right) \tag{73}$$

and $\Delta\tau_i = \chi_i \tfrac{1}{2}(z_{i+1} - z_{i-1})$ as before. The quadratic term in Eq. 72 vanishes if the line opacity in the interval is small, indicating the absence of a significant line at the next grid point. The power of 2 is necessary to ensure that this term drops to zero faster than the corresponding transport coefficient $\sim 1/\Delta\tau_i^\pm$. If a line or strong continuum is present at the current grid point, the second term becomes zero avoiding additional emission in the center of the line.

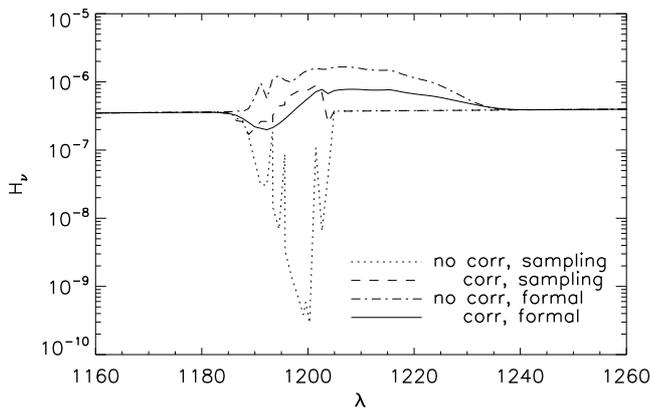

**Fig. 8.** Emergent flux in the Lyman-$\alpha$ line of a Type II Supernova model containing H and He. For better comparison, both models have been computed neglecting the contribution of these lines to the line blocking in all iterations except the last one. Thus, both models have very similar occupation numbers and temperature structures. The two lines indicated with "sampling" refer to the emergent flux in the last iteration using the opacity sampling technique. The other two lines represent the flux after the first iteration that uses the exact solution. The dotted line and the dash-dotted line belong to the model that uses the standard procedure in the solution of the Feautrier scheme; the dashed line and the solid line refer to the model that had the correction applied.

## 4. Results of test calculations

To test the properties of the new coefficients, we have simulated a variety of different situations before applying the method in the actual spectral synthesis code. To estimate the quality of the correction in realistic models, we also investigated situations with continuum opacity (in particular including Thomson-scattering). In this regard the toy model had to be extended to a spherically symmetric model for two main reasons. First, the determination of the Thomson-emissivity requires knowledge of the mean intensity $J$. Second, we intended to see the effect on the angle-integrated quantities.

Fig. 4–Fig. 7 show a selection of situations we have simulated using this extended version. All low-resolution models use a uniformly spaced radial grid with 41 grid points and 5 core rays. The high-resolution grid is set up by dividing each interval into 14 sub-intervals and uses 20 core rays. The source function of the line was again set to $S_{\text{line}} = 2$ and the incident intensity at the core to $I^+ = 4$. The opacity of the line has been determined such that a given radial optical depth $\tau_{\text{line}}$ was reached. In all figures, the upper panel shows the derived mean intensity $J$ without correction and the lower panel shows the respective models with the correction applied. All values of $J$ have been obtained by iterating the ray-by-ray solution with the solution of the moment equation. (The final $J$ from the different methods was in agreement on the level of a few percent, depending on the model.)

Fig. 4 shows the setup with an optically thick line ($\tau_{\text{line}} = 50$) and a low constant background source function and opacity ($S_{\text{bg}} = 0.1, \chi_{\text{bg}} = 0.1$). In the presence of opacity on other grid points, the uncorrected solution outside the line couples more strongly to the local source function, since the coupling to the line is too small.

In Fig. 5 we show the same situation but include Thomson scattering instead as background opacity, whereas in Fig. 6 the background consists of both true and Thomson opacity. In both these cases the solution of the corrected method in the low-resolution case is very close to the high-resolution result, with only a fraction of the depth points needed. Finally in Fig. 7, we show the case of an optically thin line ($\tau_{\text{line}} = 0.5$). In this case the new corrected and the old uncorrected method, as expected, give very similar results.

Fig. 8 shows the application of the non-LTE radiative transfer code on a test model for a SN II containing only H and He. For better comparison, both models have been computed neglecting the contribution of these lines to the line blocking in all iterations except the last one. Thus, both models have very similar occupation numbers and temperature structures. Shown in this figure is the emergent flux in the Lyman-$\alpha$ line for two models before and after the first iteration of the detailed solution. The dotted and the dash-dotted line refer to the model that has been calculated using the standard Feautrier scheme described in Section 2 without correction. The dotted line shows the flux in the last sampling iteration (method I); the dashed-dotted line represents the flux after the first iteration of the detailed solution (method II). The dashed line (method I) and the solid line (method II) represent the flux resulting from the model where the correction has been applied. One can see that the sampling iteration without the correction produces significantly less flux, which also affects the occupation numbers of the corresponding level. This also leads to spurious results in the detailed solution. With the correction included, the flux agrees much better with the detailed solution of method II (exact agreement is not possible since our sampling method considers the Doppler-shifts of the central ray as being representative for the other rays as well, cf. Pauldrach et al. 2001).

Although the error introduced is less obvious for weaker lines that are blended with other lines, a multitude of extremely strong lines (like the Lyman-$\alpha$ line in this SN II model) are to be found in SN Ia, and the standard procedure *systematically* underestimates the radiation field outside of these lines. We note, however, that the total influence on a model is not so easy to predict as in the test cases where we have held the occupation numbers fixed, since due to the changed radiation field the ionization and excitation rates may change, leading to different occupation numbers and thus different opacities and source functions of the lines.

## 5. Application to synthetic SN Ia spectra

Fig. 9 shows the application of our current non-LTE radiative transfer procedure to a SN Ia model at an epoch of 25 days ($\sim 5$ days after maximum). Besides the high expansion velocities, the fact that the atmospheres of SN Ia have no appreciable true continuum over a significant spectral range and that the pseudo-continuum is actually formed by thousands of overlapping lines that influence each other, required not only a modified Feautrier scheme, but also a solution of one of the biggest challenges we have found in the modeling of the radiative transfer in SN Ia: the transfer of the radiative energy from the UV into the optical regime via only spectral lines.

That radiation has to be transferred from the UV into the optical regime becomes evident from a comparison of the blue curve and the green curve in Fig. 9. The blue curve has its maximum in the UV spectral range, and shows the shape of the con-tinuous radiation resulting from the true continuum plus Thomson scattering opacities and emissivities of the innermost regions, whereas the green curve shows the true emergent spectrum of the model, and has most of the flux in the optical range.

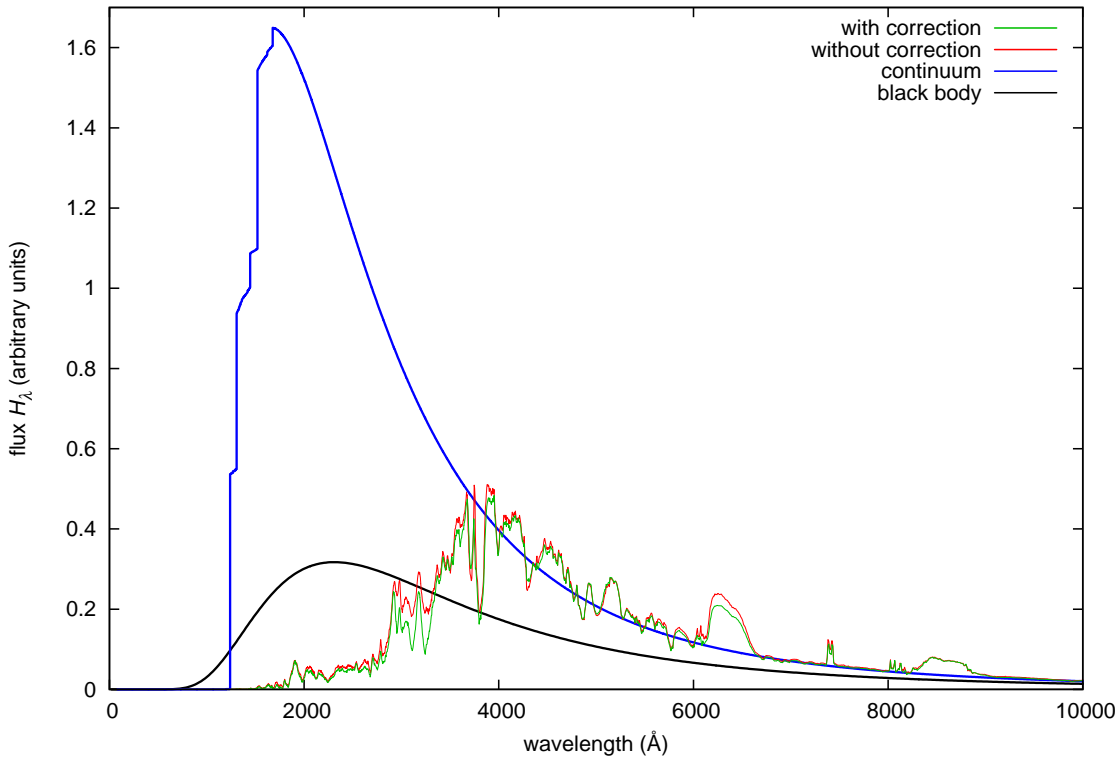

**Fig. 9.** Synthetic SN Ia spectra of consistent ALImI-models (SN Ia models which are based on an accelerated lambda iteration procedure for the mutual interaction of strong spectral lines, cf. Pauldrach et al. 2013) using the original (red curve) and the new treatment (green curve) of the Feautrier method in the first iteration cycle (shown is the flux after the first iteration that uses the detailed solution). Although the two models are not strictly comparable since the occupation numbers, the ionization, and the temperature structure adjust differently because of the differences of the Feautrier schemes applied, it is readily seen that the behavior of the flux is different especially for wavelengths which are dominated by strong lines. For comparison an equivalent blackbody spectrum (black curve) corresponding to the effective temperature of the atmosphere ($T_{\text{eff}}$ = 12 540 K) and the spectrum resulting from a formal solution without lines (blue curve) are also shown, demonstrating the very different slopes these two curves exhibit even in the wavelength range redward of 4000 Å where the supernova ejecta are much less dominated by spectral lines. (Blueward of 3000 Å the flux is of course blocked and strongly attenuated by spectral lines; flux conservation results in enhanced emission redward of 3000 Å compared to a black body.)

In most stellar objects it is the true continuous absorption and emission processes which cause this shifting of the spectral distribution from higher to lower energies, but this mechanism cannot work in SN Ia envelopes. Because of the small absolute densities (compared to stars) and a composition which is dominated by low ionization stages of intermediate-mass and iron-group elements, the free-free and bound-free continuum in the UV and optical regime is very weak in the outer parts of the atmospheres of these objects. In fact, electron scattering is the dominant source of opacity at wavelengths redward of about 5000 Å, even for deeper layers of the atmospheres. Blueward of about 4000 Å, however, the total continuous opacity is completely irrelevant when compared to the line opacities (cf. Pauldrach et al. 2013). This behavior illustrates the formation of the pseudo-continuum in the envelopes of SN Ia, which results solely from the overlap of several ten-thousands of spectral lines and which is entirely responsible for the transfer of the radiative energy from the UV into the optical regime.[7]

As the conversion of radiation through the overlap of strong lines of different ionization stages and elements does not happen when using the standard single-line accelerated lambda iteration procedure,[8] we had to develop an improved treatment ("ALImI"; cf. Pauldrach et al. 2013) for these lines that mutually interact with each other in the pseudo-continuum. Thus, hydrodynamic explosion models and realistic model atmospheres that take into account the strong deviation from local thermodynamic equilibrium are not the only ingredients necessary for the spectral synthesis and analysis of SN Ia spectra, but also improved iteration schemes and radiative transfer methods.

---

[7] Because of the large velocity gradients, which – in a state of homologous expansion with velocities reaching up to 20 000 km/s – bring photons emitted at the photosphere and redshifted on their way outwards through the envelope into resonance with redder and redder lines, the UV part of the spectrum is effectively blocked by the superposition of strong metal absorption lines The number of such lines is very large and in the UV, where the metal lines crowd very densely, this line overlap in practice becomes a continuous source of opacity (cf. Fig. 10 of Pauldrach et al. 2013). A UV photon can thus only escape if during its random walk being scattered and re-scattered in the optically thick atmosphere it manages to connect to the thermal pool, is reborn as a redder photon, and finds a low-opacity window in the spectrum (cf. Pauldrach et al. 1996).

[8] The classical accelerated lambda iteration (ALI) simply fails because it is based on the premise that only the one dominant process that has "frozen" the iteration must be canceled to let the system converge at each frequency, but in SN Ia at any frequency and depth point the radiation field is actually influenced by several tens of strong lines.

Fig. 9 further shows that the excess flux in the red and infrared wavelength regions that most of the synthetic spectra for SN Ia at early epochs calculated up to now (cf. Pauldrach et al. 1996, Nugent et al. 1997, Sauer et al. 2006, and Stehle et al. 2005) had shown compared to observations does not appear in our model. This is because our new iteration scheme allows us to calculate much deeper into the "photosphere" while still permitting the iteration to converge despite the very high optical depths encountered in the UV.

Thus the observed steep slope in the continuum in between 4000 Å and 8000 Å can be understood to be the result of a much bluer, non-Planckian radiation field in the inner regions of the ejecta, which is mostly unobservable due to the strong line blocking in the UV, but which can be inferred from the red part of the spectrum where the opacities are low. Radiative transfer models which assume a customary diffusion approximation as inner boundary farther out in the ejecta will instead show much shallower slopes in this range comparable to that of a black body corresponding to the effective temperatures of SN Ia atmospheres.

## 6. Summary and conclusions

Our investigations lead to an improved understanding of how the emergent spectra of type Ia supernovae are being shaped within the ejecta, and how this relates to the particular physical properties of their envelopes and the resulting peculiarities in their radiative transfer compared to the much better studied atmospheres of stars. As these spectra are assembled in a complex way by numerous mostly unobservable spectral features, this knowledge provides an important insight for the process of extracting information from the observed SN Ia spectra.

In this paper we have pointed out and investigated a necessary step required for a quantitative analysis of SNIa spectra based on an elaborate approach for expanding atmospheres characterized by a sophisticated treatment of the strong spectral lines which mutually interact with each other via a "pseudo-continuum" entirely formed by these Doppler-shifted spectral lines themselves, and characterized by the treatment of a consistent solution of the full non-LTE rate equations along with a detailed solution of the radiative transfer.

In particular we have shown that the Feautrier method of solving the radiative transfer equation can be extended to cases – strongly varying opacities and source functions from grid point to grid point – which by their very nature would argue against the Feautrier scheme as a solution method of choice. Indeed, the standard formulation of the Feautrier method as a finite difference method attempting to approximate the differential equation of radiative transfer fails badly in these cases. However, knowing the nature of the physical conditions underlying these cases has allowed us to construct correction factors that mimic the true behavior of the radiation field without changing the structure of the algorithm, and thus allowing for a direct plug-in replacement in the solver. Although some differences remain between the corrected Feautrier solution and the exact solution, the fact that our corrected method yields much better results than the standard formulation is of usefulness for the application of this widely employed method. In our model atmosphere code it allows us to save computer time in the "pre-iteration" step as it provides a better starting point for the final iterations using the computationally much more expensive detailed radiative transfer solver. The results presented indicate that our presently applied method is now on a level where it may be considered for use in quantitative analyses.

*Acknowledgements.* This work was supported by the Deutsche Forschungsgemeinschaft under Grant Pa 477/7-1.